\documentclass[a4paper]{amsart}

\usepackage[T1]{fontenc}
\usepackage[utf8]{inputenc}
\usepackage{amsmath}
\usepackage{amsfonts}
\usepackage{amssymb}
\usepackage{amsthm}
\usepackage{mathtools}
\usepackage[scaled]{helvet}
\usepackage{mathpazo}
\usepackage[scr=rsfso,scrscaled=1.1]{mathalpha}
\usepackage{comment}

\usepackage{graphicx}
\usepackage{tikz}
\usetikzlibrary{decorations.pathmorphing}
\usetikzlibrary{backgrounds}
\usepackage{hyperref}
\hypersetup{linkcolor=black,
            urlcolor=blue,  
            breaklinks=true,
            colorlinks=true,
            citebordercolor=0 0 0,
            filebordercolor=0 0 0,
            linkbordercolor=0 0 0,
            menubordercolor=0 0 0,
            urlbordercolor=0 0 0
} 

\usepackage[style=phys,sorting=nyt,
                articletitle=true,
                biblabel=brackets,
                chaptertitle=false,
                pageranges=false,
                hyperref=true,
                maxcitenames=3,
                url=true,
                autocite=superscript] {biblatex}

\newcommand*{\cO}{{\mathcal O}} 
\newcommand*{\cT}{{\mathcal T}} 
\newcommand*{\sS}{{\mathscr S}} 
\newcommand*{\sN}{{\mathscr N}} 
\newcommand*{\sM}{{\mathscr M}} 
\newcommand*{\scri}{\mathscr{I}}

\newcommand*{\dd}{\mathrm{d}}
\newcommand*{\del}{\partial}

\renewcommand*{\c}[1]{\mathbf{#1}}
\newcommand*{\eb}{\mathbf{e}}
\newcommand*{\hnabla}{\widehat{\nabla}}
\newcommand*{\hGamma}{\widehat{\Gamma}}
\newcommand*{\hP}{\widehat{P}}
\newcommand*{\Y}[2]{{}_{#1}Y_{#2}}

\newcommand*{\new}[1]{#1}

\begin{document}

\title{A numerical framework for studying asymptotic quantities}

\author[B. Camden]{Breanna Camden}
\email{breanna.camden@pg.canterbury.ac.nz}
\address{School of Mathematics and Statistics,
  University of Canterbury, Christchurch 8041, New Zealand}

\author[J. Frauendiener]{Jörg Frauendiener}
\thanks{JF would like to thank the Isaac Newton Institute for
Mathematical Sciences, Cambridge, for support and hospitality during the programme ”Twistor theory” where part of the work on this paper was undertaken. This work was supported by EPSRC grant no EP/R014604/1 and by the Marsden Fund Council from Government funding, managed by Royal Society Te Ap\={a}rangi.}
\email{joerg.frauendiener@otago.ac.nz}
\address{Department of Mathematics and Statistics, University of Otago, New Zealand}

\author[J. Galinski]{Joseph Galinski}
\email{joseph.galinski@pg.canterbury.ac.nz}
\address{School of Mathematics and Statistics,
  University of Canterbury, Christchurch 8041, New Zealand}

\author[K. Pillay]{Kaushal Pillay}
\email{pillaykaushal@gmail.com}
\address{Department of Mathematics and Statistics, University of Otago, New Zealand}

\author[C. Stevens]{Chris Stevens}
\email{chris.stevens@canterbury.ac.nz}
\address{School of Mathematics and Statistics,
  University of Canterbury, Christchurch 8041, New Zealand}

\author[S. Thwala]{Sebenele Thwala}
\email{sebenele.thwala@pg.canterbury.ac.nz}
\address{School of Mathematics and Statistics,
  University of Canterbury, Christchurch 8041, New Zealand}
\begin{abstract}
  In this contribution we present \new{an overview of} our work on the numerical simulation of the perturbation of a black hole space-time by incoming gravitational waves. The formulation we use is based on Friedrich's general conformal equations which have the unique property that they allow access to the asymptotic region of an asymptotically regular space-time. In our approach we set up an initial boundary value problem on a finite boundary, which cleanly separates the initial conditions, a static black hole, from the perturbation, an incoming gravitational wave specified by a spin-2 function on the time-like boundary. The main advantage of this approach is that the finite boundary expands fast enough to reach null-infinity where the asymptotic properties can be studied. This provides, for the first time, a direct relationship between finite initial and boundary data and asymptotic quantities within one simulation. We discuss the possibilities and limitations of this approach.
\end{abstract}
\maketitle

\section{Introduction}
\label{sec:introduction}

A large effort of numerical relativity is spent on the computation of waveforms, the gravitational radiation signals emitted by astrophysical processes in the Universe. The main tool for this computation is the idea of an isolated system. This is the idealisation of a physical system (a star, a galaxy, a galaxy cluster) which evolves under its own gravitational field and has only weak interactions with its environment, see~\cite{Geroch:1977} for an illuminating discussion of this issue. For example, in a binary system, each constituent by itself would not be isolated individually, but the system as a whole would be.

In this idealisation, the gravitational field surrounding the system would have to decay at large distances, leading to an asymptotically vanishing curvature, i.e., an asymptotically flat space-time. This idea was made precise by casting it into a geometric framework by Penrose~\cite{Penrose:1965} and will be discussed further below.

Of course, as with every idealisation, there are limiting cases: consider two massive bodies, initially far apart, approaching each other. Two scenarios can happen: they can come to a closest distance and then separate again, or they form a bound state, ultimately colliding and merging into a single body. 

Should this case be considered as isolated? This depends to some extent on how we think of the motion of the two bodies in the infinite past. The usual description would be to think of the bodies naively as coming in "from infinity" which is an idealisation by itself. If we think in this way then, clearly, the system is not isolated since there is influence from infinite distances and from the infinite past. With the same reasoning, we would not attribute the idea of being isolated to the first case, described above, of the two bodies separating indefinitely in the future.

This naive reasoning must be refined when the geometric framework is considered. This is because in the geometric setting, the bodies will emerge from past time-like infinity and move towards future time-like infinity while the main concern of asymptotically flat space-times is null-infinity where the gravitational radiation is registered. The two cases show that there can be a difference between the future and the past behaviour of the system. In both cases, we would not regard the system as isolated in the past. However, the second case could be argued to be isolated in the future while the first case can not be regarded as isolated in the future.

This conclusion would also hold in the geometric setting where gravitational radiation is taken into account because the system radiates for all times and any very early outgoing radiation may have the effect of spoiling the fall-off behaviour of the gravitational field near past null-infinity in the same way as any late ingoing radiation spoils the fall-off near future null-infinity. The influence on the fall-off properties by early outgoing or late ingoing radiation can be ultimately traced back to the behaviour of the gravitational field near space-like infinity, and this issue has not yet been resolved satisfactorily.

It appears in various forms such as the violation of the ``peeling property'', an intriguing conspiracy between the fall-off and the algebraic properties of components of the Weyl tensor~\cite{Sachs:1960}, or, equivalently, as the (lack of) smoothness of null-infinity considered as a boundary to space-time (see~\cite{Kehrberger:2024}), or, in the question as to what are the necessary and sufficient conditions to impose on initial data which guarantee that no essential physical situation is being excluded. This question of ``peeling or not peeling'' has been discussed at length by Friedrich~\cite{Friedrich:2018}. His method of analysing the structure at space-like infinity in the form of a finite regular initial boundary value problem~\cite{Friedrich:1998} is beyond the scope of the present discussion but it is quite fundamental for the resolution of several questions about structures at infinity.

The fact that space-like infinity is still in some sense ``terra incognita'' is also reflected in the fact that there are no numerical treatments of the Einstein equations which include this region. Most codes avoid it by introducing an artificial outer boundary, while others start the evolution from a spatially infinite initial surface, whose conformal boundary is in the future of space-like infinity. The situation described in the present paper interpolates between these two approaches in the sense that initial data are given on a bounded space-like hypersurface which over the course of the evolution expands until it reaches and intersects null-infinity, thereby becoming a hyperboloidal hypersurface.

The paper is structured as follows. In section~\ref{sec:gener-conf-field} we talk about the mathematical background, such as the general conformal field equations and the conformal Gauß gauge. Section~\ref{sec:more-deta-form} provides more specific details of the formulation and presents a short discussion of our numerical implementation. In section~\ref{sec:some-results} we discuss the results which we have obtained so far and in the final section we look at the possibilities and limitations that come with our approach. \new{Since this paper is intended as giving an overview of this work we do not provide detailed convergence tests and other justifications but refer to published articles instead.} The conventions used in this paper are those of \new{Penrose and Rindler}~\cite{Penrose:1984a}.

\section{The general conformal field equations and the conformal Gauß gauge}
\label{sec:gener-conf-field}

The unique feature of our approach is the fact that the initially finite time-slices expand during the evolution until they intersect null-infinity, forming hyperboloidal hypersurfaces. From the first instant, when this intersection happens, we are dealing with a hyperboloidal initial value problem.

Our formalism is based on Friedrich's~\cite{Friedrich:1995} generalised conformal field equations (GCFE), a set of geometric PDEs which formulates the fact that a conformal class $[g_{ab}]$ of metrics on a space-time $\sM$ contains a physical vacuum metric $\tilde{g}_{ab}$. By construction, the formalism is conformally invariant in the sense that the vacuum metric is represented in terms of a metric $g_{ab}$ in the conformal class and a conformal factor $\Theta$, such that 
\[
  g_{ab} = \Theta^2\tilde{g}_{ab},
\]
which entails the gauge freedom $(\Theta,g_{ab})\mapsto (\omega \Theta,\omega^2\,g_{ab})$ for some positive function $\omega$. The full system of equations consists of Cartan's structure equations for a so-called Weyl connection, two equations for the curvature which are derived from the Bianchi identities for both the physical and conformal Riemann tensors, and the Einstein vacuum equations for $\tilde{g}_{ab}$. All quantities are expressed with respect to a coordinate system $(x^\mu)_{\mu = 0:3}$, a frame $(\eb_{\c{a}})_{\c{a}=0:3}$ with frame coefficients $c^\mu_{\c{a}}$ defined by $\eb_{\c{a}}=c^\mu_{\c{a}}\,\del_\mu$, which defines a metric $g_{ab}$ in the conformal class, and a Weyl connection $\hnabla_a$ with connection coefficients $\hGamma^{\c{c}}_{\c{ab}}$ with respect to that frame. Using these choices, the equations become
\begin{align}
    \eb_{\c{a}}(c^\mu_{\c{b}})  -  \eb_{\c{b}}(c^\mu_{\c{a}}) &= \hGamma^{\c{c}}_{\c{ab}}c^\mu_{\c{c}} - \hGamma^{\c{c}}_{\c{ba}}c^\mu_{\c{c}},\label{eq:1}\\
    \eb_{\c{a}}(\hGamma^{\c{d}}_{\c{bc}}) -  \eb_{\c{b}}(\hGamma^{\c{d}}_{\c{ac}} ) &= \hGamma^{\c{e}}_{\c{ab}}\hGamma^{\c{d}}_{\c{ec}} - 
    \hGamma^{\c{e}}_{\c{ba}}\hGamma^{\c{d}}_{\c{ec}} -
\hGamma^{\c{e}}_{\c{bc}}\hGamma^{\c{d}}_{\c{ae}} +
\hGamma^{\c{e}}_{\c{ac}}\hGamma^{\c{d}}_{\c{be}} \nonumber \\
&+ \Theta K_{\c{abc}}{}^{\c{d}} - 2 \eta_{\c{c}[\c{a}} \hP_{\c{b}]}{}^{\c{d}} + 2 \delta_{[\c{a}}{}^{\c{d}}\hP_{\c{b}]\c{c}} - 2 \hP_{[\c{a}\c{b}]}\delta_{\c{c}}{}^{\c{d}},\label{eq:2}\\
\hnabla_a\hP_{bc} - \hnabla_b\hP_{ac} &= \Theta b_e K_{abc}{}^e,\label{eq:3}\\ 
    \nabla_e K_{abc}{}^e &= 0.\label{eq:4}
\end{align}
In these equations, the tensor $K_{abc}{}^d := \Theta^{-1}C_{abc}{}^d$ is the rescaled Weyl tensor, describing the gravitational field. The Weyl form $b_a$ is defined by the operation of the Weyl connection on the physical metric: $\hnabla_c\tilde{g}_{ab}=-2b_c \tilde{g}_{ab}$. The tensor $\hP_{ab}$ is the Rho- or Schouten tensor (see~\cite{Penrose:1984a})\footnote{\new{Note, that, even though related, this tensor is not identical to Geroch's tensor $\rho_{ab}$, as defined in~\cite{Geroch:1977}}.} of the Weyl connection and $\nabla_a$ is the Levi-Civita connection of the metric $g_{ab}$.

This is a set of geometric PDEs in the sense that they fix a geometric structure, here the physical vacuum space-time defined by the metric $\tilde{g}_{ab}$, but not the representation of that structure with respect to a specific coordinate system and other arbitrary choices. This is reflected in the large gauge freedom that is present in these equations. As we have seen, we needed to choose the Weyl connection, the conformal factor, the frame and the coordinate system in order to characterize the desired vacuum metric. The arbitrariness of this choice is reflected in the explicit appearance of $\Theta$ and $b_a$ for which there are no equations, and by the well-known underdetermined nature of the structure equations~\eqref{eq:1} and~\eqref{eq:2}.

Fixing these quantities is conveniently achieved by the so-called conformal Gauß gauge (CGG), introduced by Friedrich~\cite{Friedrich:1995}, and further discussed in the context of our framework in~\cite{Beyer:2017}. It is based on a congruence of time-like conformal geodesics and fixes all these choices at once. A remarkable fact of this gauge is that the conformal factor $\Theta$ can be determined explicitly in terms of initial data with respect to the simultaneously fixed coordinate system.

Imposing the CGG has several remarkable consequences. The first one is global and it is the main reason for the feasibility of our approach. Friedrich~\cite{Friedrich:2003} analysed the behaviour of time-like conformal geodesics in the Schwarzschild space-time and found that these curves cover the entire Kruskal-Szekeres extension. In our application we study the impact of gravitational waves on a Schwarzschild black hole and we expect that this will result in a deformed Schwarzschild space-time. Thus, it is not unreasonable to assume that a similar construction of a congruence of conformal time-like geodesics on the deformed space-time will have similar properties, so that the imposition of the CGG will make sense; and this is indeed exactly what we find.

The second consequence is of practical importance and has to do with the complexity of the equations when written in that gauge. In the same way as other formulations of the Einstein equations, once a time coordinate has been singled out, the GCFE can be formally split into constraint equations and evolution equations. Of these two subsystems the constraints are uniquely determined while the evolution equations are generally fixed only ``modulo the constraints''.

Now it turns out, that when expressed in the conformal Gauß gauge, the evolution equations simplify significantly. Of all quantities, only the components of the rescaled Weyl tensor satisfy a symmetric hyperbolic system while all others propagate along the time direction, i.e., they satisfy simple advection equations. This shows explicitly that the gravitational wave degrees of freedom are carried by the Weyl tensor.

Our application is based on this example. Instead of a global initial data hypersurface which is asymptotically Euclidean we use a space-like hypersurface with (finite) boundary. Under the evolution with the GCFE in the conformal Gauß gauge, the boundary is dragged along the congruence of conformal geodesics, forming a time-like hypersurface on which boundary data must be prescribed in order to obtain a determined system. Due to the properties of the conformal geodesics, the time-like boundary expands so rapidly that it ultimately intersects null-infinity in a 2-dimensional surface, a cut. From this ``instant'' onwards, we have future null-infinity $\scri^+$ inside our computational domain and we can analyse the properties of the physical quantities defined there. The hypersurfaces of constant time expand in a corresponding way and, once they intersect null-infinity, become hyperboloidal. In this sense, our approach dynamically generates hyperboloidal hypersurfaces from bounded space-like hypersurfaces. Thus, we determine hyperboloidal initial data sets without solving any elliptic equations.

The structure of the evolution equations has the consequence that most quantities propagate along the boundary while only the gravitational waves represented by the rescaled Weyl tensor travel across the boundary. This allows us to cleanly separate the initial state from the exterior influence, the ingoing radiation. In our application, we initially specify data for an exact (static) Schwarzschild black hole, and we direct gravitational radiation from the outer boundary onto this system to drive it away from staticity.

\section{More details on the formulation and the numerical setup}
\label{sec:more-deta-form}

\subsection{The conformal Gauß gauge}
\label{sec:conformal-gau3-gauge}

The conformal Gauß gauge is imposed by the following conditions (see~\cite{Beyer:2017} for details):
\begin{itemize}
\item The coordinates are defined by dragging the spatial coordinates along the conformal geodesics, which are therefore given by equations of the form $(x^1,x^2,x^3) = const$. The time coordinate $t$ is taken to be the preferred parameter along the curves.
\item The Weyl connection $\hnabla_a$ is fixed by the additional initial conditions for the 1-form $b_a$ defined by the conformal geodesic equations. 
\item A frame  $\eb_{\c{a}}$, orthogonal with respect to the physical metric $\tilde{g}_{ab}$, is defined by initially choosing it to be adapted to the initial hypersurface, i.e., $\eb_{\c{0}}$ is perpendicular to $\Sigma$ and the spatial frame vectors are tangent to it. The frame is extended into $\sM$ by propagating it with the Weyl connection $\hnabla_a$. This has the consequence that the frame does not necessarily remain adapted to the hypersurfaces of constant time $t$.
\item This frame defines a metric $g_{ab}$ in the conformal class of $\tilde{g}_{ab}$ with respect to which it is orthonormal. This, in turn, defines the conformal factor $\Theta$ by the formula
  \[
    \Theta^2\tilde{g}(\eb_{\c{a}},\eb_{\c{b}}) = \eta_{\c{ab}} = g(\eb_{\c{a}},\eb_{\c{b}}).
  \]
In the conformal Gauß gauge, this function is completely determined in terms of initial data by
\begin{equation}
  \label{eq:5}
  \Theta(x^\mu) = \underline{\Theta}\left(1 + \frac{t^2}{4\eta^{\c{ab}}\underline{b}_{\c{a}}\underline{b}_{\c{b}}}\right).
\end{equation}
Here, the underlined quantities denote functions which are constant along the conformal geodesics.
\end{itemize}
As a consequence of these choices, we find that
\begin{equation}
  c_{\c{0}}^\mu = \delta_0^\mu,\quad \hGamma_{0\c{a}}^{\c{b}}=0, \qquad \hP_{\c{0a}}=0, \qquad \eb_{\c{0}}^c\hnabla_c g_{ab} = 0.\label{eq:13}
\end{equation}

\begin{figure}[h]
    \centering
\begin{tikzpicture}[very thick, decoration = {bent,amplitude=-5},scale=0.7]
  \draw[dashed,blue!40] (2,-2) -- (0,0)--(2,2);
  \draw[blue!60] (2,2)--(4,0) -- (2,-2);
  \draw[dashed,blue!40] (-2,-2) -- (0,0)--(-2,2);
  \draw[blue!60] (-2,-2)--(-4,0) -- (-2,2);
  \draw[red,decoration={bumps,amplitude=-2},decorate] (-2,2) .. controls (0,1.8) .. (2,2);
  \draw[red,decoration={bumps,amplitude=2},decorate] (-2,-2)  .. controls (0,-1.8) .. (2,-2);
  \draw[draw  = green, very thick, fill=green!20!white, fill opacity=0.3] 
                          (0,1.8) -- (0,0) -- (3.5,0) -- (3.5,1.8);
\draw[green, thin] (0,0) -- (0,1.8);
\foreach \x in {0.0, 0.5,...,3.5} \draw[green, thin] (\x,0) -- (\x,2.0);
\node at (4,0)[anchor=west]{$i^0$};
\node at (2,2)[anchor=south]{$i^+$};
\node at (2,0)[anchor=north]{$\Sigma$};
\node at (3.5,1)[anchor=west]{$\cT$};
\end{tikzpicture}
\caption{A schematic diagram showing the computational domain, shaded in green, in relation to the Kruskal extension of the Schwarzschild space-time. The vertical thick green lines are the inner and outer boundaries. On the horizontal green line we specify initial data. The other green lines indicate the time lines. Note, that we stay away from space-like infinity~$i^0$. Also note, that at $i^+$ the horizon, the singularity and null-infinity approach each other.}
      \label{fig:domain}
\end{figure}
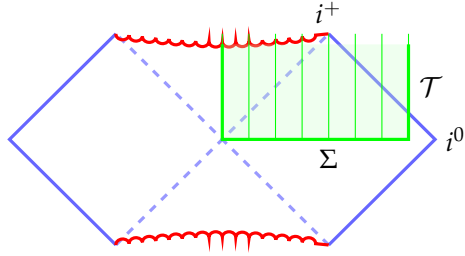

\subsection{The evolution equations}
\label{sec:evolution-equations}

Inserting these expressions into~(\ref{eq:1}-\ref{eq:4}) we can split these equations into evolution and constraint equations according to whether they contain derivatives in the direction of $\eb_{\c0}$ or not. This yields the system of evolution equations 
\begin{align}
    \eb_{\c{0}}(c^\mu_{\c{b}})   &=  - \hGamma^{\c{c}}_{\c{b0}}c^\mu_{\c{c}},\\
    \eb_{\c{0}}(\hGamma^{\c{d}}_{\c{bc}})  &= - \hGamma^{\c{e}}_{\c{b0}}\hGamma^{\c{d}}_{\c{ec}} + \Theta K_{\c{0bc}}{}^{\c{d}} -  \eta_{\c{c}\c{0}} \hP_{\c{b}}{}^{\c{d}} + \delta_{\c{0}}{}^{\c{d}}\hP_{\c{b}\c{c}} + \hP_{\c{b}\c{0}}\delta_{\c{c}}{}^{\c{d}},\\
\eb_{\c{0}}(\hP_{\c{bc}}) &= -\hGamma^{\c{e}}_{\c{b0}} \hP_{\c{ec}} + \Theta b_{\c{e}} K_{\c{0bc}}{}^{\c{e}},\\ 
    \eb_{\c{0}}(K_{\alpha}) &= \mathbf{A}_{\alpha\beta}^{\c{i}}\eb_{\c{i}}( K^\beta) + \pmb{\hGamma}_{\alpha\beta} K^\beta.\label{eq:14}
\end{align}
Here, we left the subsystem~\eqref{eq:14} for the rescaled Weyl tensor in symbolic form; the indices $\alpha$ and $\beta$ are ``clumped indices'', indicating various components of $K_{abc}{}^d$, on which the symmetric matrices $\left(\mathbf{A}^{\c{i}}\right)_{\c{i}=1:3}$, and $\pmb{\hGamma}$ operate. It can be written most naturally in terms of the electric and magnetic parts of $K_{abc}{}^d$. In this form, it is quite straightforward to show that the subsystem is symmetric hyperbolic. Its explicit form can be found in~\cite{Beyer:2017,Frauendiener:2004a} and references therein.

We do not exhibit the constraint equations explicitly here, mainly for the reason that they do not play a big role in this discussion. However, we should remark that the constraints and evolution equations  are compatible in the sense that solutions of the constraints evolve into solutions of the constraints. Therefore, the constraints need to be satisfied only on the initial hypersurface. We achieve this by computing the relevant geometric quantities from the exact solution of a black hole so that the constraints are automatically satisfied on the initial hypersurface.

\subsection{The boundary}
\label{sec:boundary}

The initial hypersurface $\Sigma$ is space-like and assumed to have two 2-dimensional spherical boundaries, the inner and the outer boundary, see Fig.~\ref{fig:domain}. In the following, we will focus on the outer boundary, labeled $\sS$. In principle, the same discussion holds at the inner boundary. During the evolution, $\sS$ is dragged along the conformal geodesics, thereby generating a time-like tube $\cT$, which is guaranteed to be a regular submanifold at least for some time into the future. Since the gravitational field $K_{abc}{}^d$ satisfies a symmetric hyperbolic system, it has characteristic cones which means that its components can propagate across the boundary. There are three characteristic cones, the light-cone, a time-like cone and a degenerate cone which coincides with the time lines. Therefore, the components crossing into the computational domain---in an appropriately adapted null tetrad they would be denoted by $\Psi_0$ and $\Psi_1$---need boundary conditions, while the outward crossing components, $\Psi_3$ and $\Psi_4$, are left alone. The remaining component, $\Psi_2$, propagates along the boundary and is entirely determined by its initial condition on~$\sS$.

This analysis of the evolution system for $K_{abc}{}^d$ implies that there are two ingoing modes while, on physical grounds, one would expect that there is only one degree of freedom at the boundary, given by $\Psi_0$ which propagates along the light-cone, i.e., with the speed of light.

The resolution of this discrepancy lies in the ``subsidiary system'', which regulates the propagation of the constraint quantities, i.e., those quantities whose vanishing is equivalent to the constraints being satisfied. This system is also symmetric hyperbolic, consisting of three complex equations for three complex constraint quantities. The characteristic cones of this system are time-like, one being the degenerate cone along the time lines and the other coinciding with the time-like cone of the evolution equations. Thus, on the boundary, one of the characteristic quantities propagates inwards, one outwards and one along the boundary.  On $\Sigma$, all three quantities vanish because the constraints are satisfied initially. Hence, imposing the boundary condition that the inward moving component vanishes on the boundary~$\cT$, so that there are ``no incoming constraint violating modes'', yields a relationship between $\Psi_0$ and $\Psi_1$, which can be used to express $\Psi_1$ in terms of $\Psi_0$. This leaves the full freedom to specify $\Psi_0$ and completely removes the additional degree of freedom in $\Psi_1$.

The remaining degree of freedom, represented by $\Psi_0$, corresponds to gravitational radiation coming in from all directions, with arbitrary angular distribution and arbitrary wave profile (duration, frequency, polarisation). \new{Within the framework of maximally dissipative boundary conditions which we use here, the ingoing wave can be made to depend on the outgoing wave, thereby modeling reflexive and transmissive properties of the boundary. Since this would attribute material properties to the boundary which are most likely unphysical, we do not pursue this further here (see, however, a plane wave example~\cite{Frauendiener:2014b}, where such boundary conditions have been explored)}.

It is worth mentioning here that a more accurate boundary condition would have to take the non-linear back-scattering into account. This would be an issue which is very complicated to resolve because it would have to take the entire history of the system along the boundary into account. There are mathematical formulations available for such totally transparent boundary conditions but we have not explored these in any detail.

\subsection{The numerical implementation}
\label{sec:numer-impl}

We have developed a general Python framework, COFFEE, for solving evolution problems, particularly the initial boundary value problem (IBVP) for the conformal field equations. This code is in the public domain and presented in detail in~\cite{Doulis:2019}. It allows for the use of general grids and discretisation methods while the time evolution is implemented by the method of lines.

For the treatment of the IBVP described here, we used a hybrid method, where the spatial domain is foliated by spherical 2-surfaces. This is general enough to allow for the simulation of nearly spherical or axisymmetric systems. On each spherical leaf we use a pseudo-spectral method based on spin-weighted spherical harmonics $\Y{s}{lm}$ and the associated ``eth'' calculus~\cite{Eastwood:1982,Goldberg:1967} as described in~\cite{Beyer:2015,Beyer:2016}. This choice makes it easy to switch between axisymmetric and general situations by simply restricting the expansion into spherical harmonics to the case $m=0$.

Across the leaves, i.e., for discretisations along the radial direction we use a $4^{\text{th}}$ order spatial finite difference method. At the outer boundary, we use summation by parts as discussed in~\cite{Gustafsson:1979,Gustafsson:1995,Kreiss:1974,Strand:1994,Sarbach:2011} together with the SAT method of Carpenter et~al.~\cite{Carpenter:1994,Carpenter:1993a} in order to implement the boundary condition for the incoming fields.

The code is parallelised in the sense that the computational domain can be split into a number of ``annuli'', each handed over to a different process. The different chunks need to exchange information needed for the radial finite difference discretisation. We implement this using ghost zones and interchanging the information using MPI. Another possibility to achieve this would have been to implement the SAT method also across the domain interfaces. However, this is computationally costly since it requires the transformation to characteristic quantities at every interface. Thus, we decided to minimize computational cost at the expense of increased inter-process communication.

\section{Some results and outstanding problems}
\label{sec:some-results}

We have looked at various aspects of the following scenario. Consider a spherically symmetric, static black hole as described by the Schwarzschild solution. Choose a sphere $\sS$ concentric with and surrounding the black hole, serving as the outer boundary of the computational domain. Another concentric sphere inside the horizon serves as the inner boundary. The time-like world-tube which is described by the time-like conformal geodesics starting at the inner boundary will always remain inside the horizon so that each of its intersections with hypersurfaces of constant time will be trapped. This means that there is no incoming information from this boundary and we do not need to perform the boundary treatment there. Instead, at the points of the inner boundary we impose the equations using a one-sided $4^{\text{th}}$ order finite difference operator to compute the spatial derivatives. This procedure results in a stable numerical evolution.

Assume that from some instant of time, gravitational radiation crosses into the computational domain from the outer boundary, forcing the black hole to react and thereby driving it away from staticity. The reaction of the black hole generates gravitational waves which propagate outwards until they can ultimately be measured on~$\scri^+$. The goal of this ``experiment'' was to obtain a qualitative and quantitative understanding of the non-linear interaction of gravitational waves with the strong curvature of space-time and the relationship between the incoming and the outgoing radiation.

The initial conditions for this situation are obtained by first writing the metric $\tilde{h}_{ab}$ as induced on $\Sigma$ by the exact Schwarzschild metric in terms of isotropic coordinates, obtaining after a minor redefinition of the radial coordinate to make it dimensionless,
\begin{equation}
  \tilde{h} = - m^2\frac{(1+r)^4}{4r^2} \left(\frac{\dd r^2}{r^2} +  \dd\sigma^2\right)\label{eq:9}
\end{equation}
with $\dd\sigma^2$ being the unit-sphere metric. Note, that in this form it is easy to see that $\tilde{h}$ is invariant under the inversion $r\mapsto r^{-1}$ which interchanges the two asymptotic ends of the Kruskal extension of the Schwarzschild space-time. In these coordinates, the original Schwarzschild radial coordinate, the areal radius $R$, is given by
\begin{equation}
  \label{eq:10}
  R = m \frac{(1+r)^2}{2r},
\end{equation}
also invariant under the inversion, so that we can write the metric in the simple form
\begin{equation}
  \label{eq:11}
  \tilde{h} = - R^2 \left(\frac{\dd r^2}{r^2} +  \dd\sigma^2\right).
\end{equation}
This metric is adapted to the asymptotic end $i_1$ described by $r\to\infty$, since $R=\cO(r)$ near $i_1$ so $\tilde{h}$ becomes asymptotically Euclidean in that limit. The other asymptotic end $i_2$, is obtained for $r\to0$. In that case, we have $R=\cO(r^{-1})$ and the corresponding metric adapted to this end is obtained by rescaling $\tilde{h}$ with the conformal factor $\Omega=(m/R)^{2}$ giving
\begin{equation}
  \label{eq:12}
  h = \Omega^2 \tilde{h} = - m^4R^{-2} \left(\frac{\dd r^2}{r^2} +  \dd\sigma^2\right).
\end{equation}
This metric becomes asymptotically Euclidean for $r\to0$.

We choose this representation of the induced Schwarzschild metric on $\Sigma$ in terms of $h$ and $\Omega$ as initial data for our simulation because using $h$ exhibits $i_2$ while the information about $i_1$ is contained in $\Omega$. This choice is well adapted to the Kruskal extension and allows us to perform our computations as closely as possible to the schematic diagram given in Fig.~\ref{fig:domain}.

From these data, together with the (vanishing) extrinsic curvature of $\Sigma$ we calculate the connection coefficients, the Schouten tensor and the gravitational field tensor. The (pull-back to $\Sigma$ of the) 1-form $b_a$ is chosen to be equal to $\dd\Omega/\Omega$ on $\Sigma$ and the initial value for the conformal factor becomes $\underline{\Theta}=\Omega$. Then the conformal Gauß gauge implies
\[
  \Theta = \frac{4r^2}{(1+r)^4} - t^2 \left(\frac{r-1}{r+1}\right)^2\qquad b_{\c{0}} = \frac{\dot\Theta}{\Theta}, \qquad b_{\c{1}} = \frac{1-r^2}{r m}. 
\]
The boundary data are given on $\cT$ as a complex spin-weight $s = 2$ function $\Psi_0(t, \vartheta, \varphi)$, which we parametrise quite generally as
\begin{equation}
  \label{eq:6}
  \Psi_0(t,\vartheta, \varphi) = \sum_{lm}a_{lm}(t) \;\Y{2}{lm}(\vartheta, \varphi).
\end{equation}
These data are adapted according to the situation at hand.

\begin{figure}[htb]
  \centering
  \includegraphics[width=0.8\textwidth]{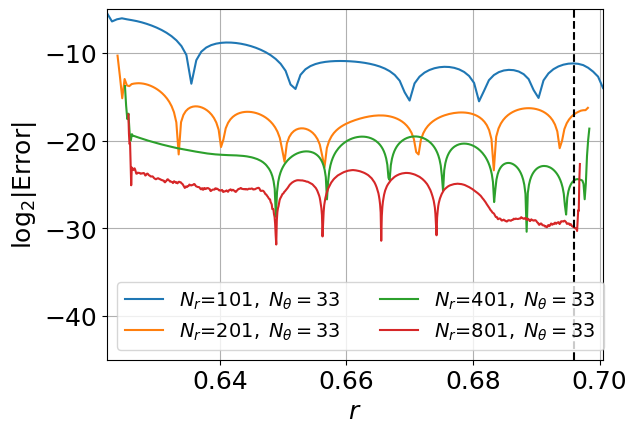}
  \caption{Convergence of the constraint violation with radial resolution at a fixed angular direction ($\theta=\pi/4$) confirming $4^{\text{th}}$ order accuracy. \new{Here, ``Error'' refers to the pointwise magnitude of the spin-0 constraint of the Bianchi system}. The vertical line represents $\scri^+$. This figure was taken from~\cite{Frauendiener:2021a}}
  \label{fig:constraints}
\end{figure}

The first tests of the code concerned its capability to reproduce the Schwarz\-schild solution and to check how well the constraint equations remain satisfied. In Fig.~\ref{fig:constraints}, we quote one of our results given in~\cite{Frauendiener:2021a}. 
We also found that the IBVP is numerically well-posed in the sense that the constraints remain satisfied and numerical errors stay bounded at the boundary.

\subsection{The Bondi-Sachs massloss}
\label{sec:bondi-mass-loss}

Our main test of the validity of the code is the check that the Bondi-Sachs massloss formula~\cite{Bondi:1962,Sachs:1962a,Newman:1962b} is valid. This is a very comprehensive test since it depends on every aspect of the code: the correct implementation of initial data, correct evolution up to null-infinity, appropriate extraction of asymptotic quantities and, finally, the correct calculation of the Bondi-Sachs energy-mo\-men\-tum and the corresponding radiation flux.

This last step is quite complicated. \new{The reason is that most of the expressions for the energy-momentum and the flux that can be found in the literature are usually obtained in a gauge which is adapted to the intrinsic structure of $\scri$. Some notable exceptions are~\cite{Geroch:1977,Ashtekar:1981,Fernandez-Alvarez:2022}. In particular, it is often assumed that the cuts of~$\scri$ on which the quantities are calculated are unit spheres.}

This assumption is far from being true in our evolution which is determined by the conformal Gauß gauge. It is a complicated transformation from the code quantities to the Bondi quantities which involves the solution of several elliptic equations on a sphere. The entire process is described in detail in~\cite{Frauendiener:2022,Frauendiener:2021a}.

\begin{figure}[htb]
  \centering
  \includegraphics[width=0.8\textwidth]{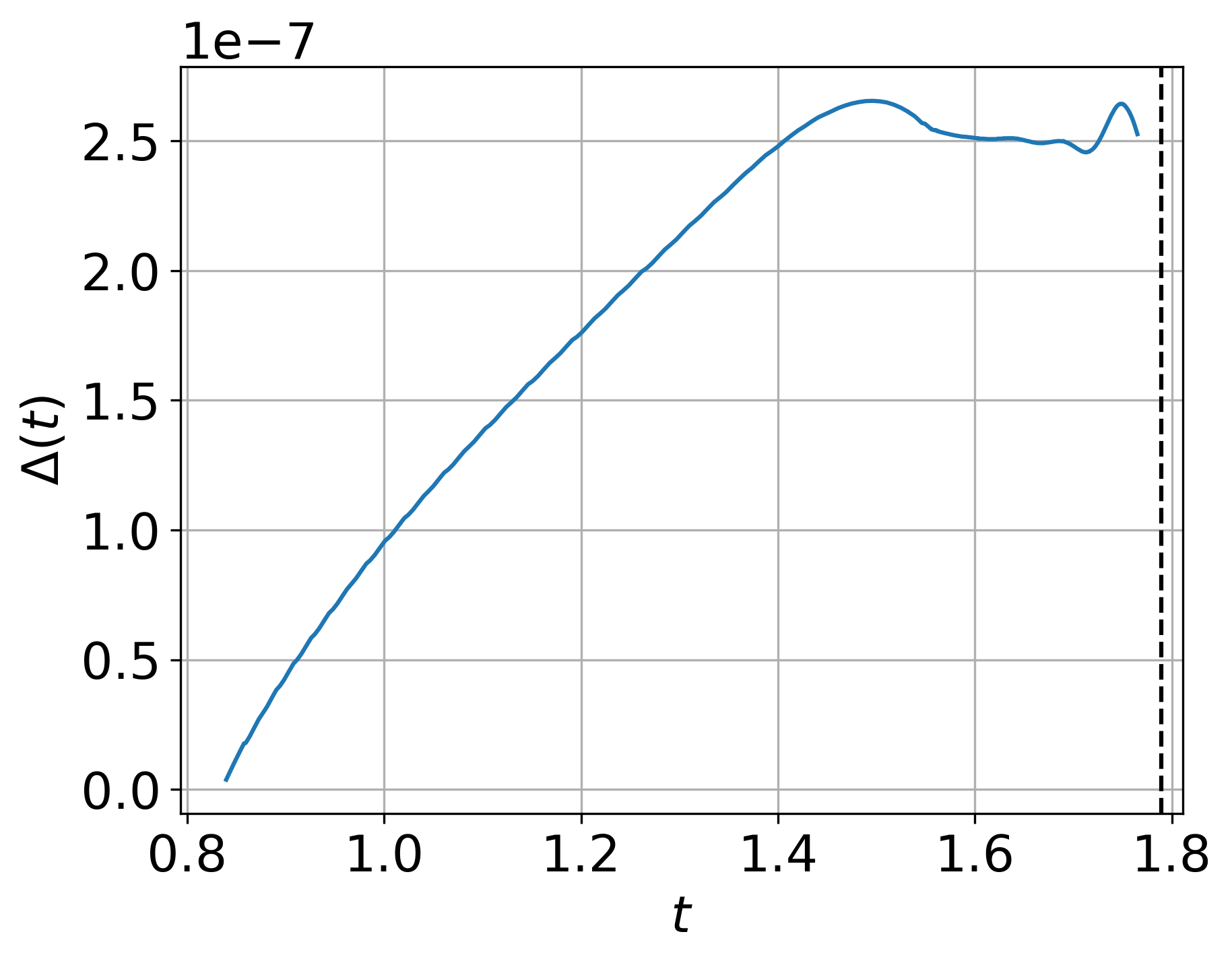}
  \caption{\new{The residual of the massloss formula relative to the Bondi mass on the first accessible cut depending on conformal time $t$}. The amplitude to produce this plot was $a=10$.}
  \label{fig:massloss}
\end{figure}

We tested the massloss by directing a single $l=2$ pulse onto the black hole, i.e., the boundary data were
\begin{equation}
  \Psi_0(t) =
  \begin{cases} 
    a\mathrm{i}\;\sqrt{\frac{2\pi}{15}}\sin^8(4{\pi t })\;\Y{2}{20}& t \leq\frac14 \\	    0 & t >\frac14
  \end{cases}, 
  \label{eq:7}
\end{equation}
where $a$ is the amplitude of the pulse.
In Fig.~\ref{fig:massloss}
we display the relative residual of the massloss, i.e., the quantity
\begin{equation}
  \Delta(t) = \frac1{m_0}\left(m_t-m_0 + \frac1{4\pi}\int_0^t\sN\overline{\sN}\,\dd^3V \right)\label{eq:8}
\end{equation}
where $m_0$ is the Bondi energy on the first cut that can be seen in the computational domain, $m_t$ is the Bondi energy on a later cut, and $\sN$ is the appropriately defined news function. The integral extends over the 3-dimensional region of $\scri$ between the two cuts. The figure shows that the massloss is numerically satisfied to a level of $10^{-7}$ \new{which agrees in order of magnitude with the constraint violation shown earlier}. The resolution used for this calculation was $801$ radial points and $33$ points in the latitudinal direction. The fact that the massloss formula is numerically valid to this level gives us confidence that the code works properly.

\new{We have also computed the Bondi mass $m_0$ on the first accessible cut depending on the incoming amplitude $a$. The result is summarised in Tab.~\ref{tab:BEs} where we give the difference between $m_0$ and the initial mass $m$ of the black hole.
\begin{table}[!h]
	\begin{center}
	\begin{tabular}{ |l|l|l|l|l| } 
	 \hline
	 $a$ & 1 & 2 & 5 & 10 \\ 
	 \hline
	 $m_0-m$ & 0.00007 & 0.00028 & 0.00173 & 0.00685 \\ 
	 \hline
	\end{tabular}
	\caption{The Bondi energy evaluated on the first cut of $\mathscr{I}^+$ for different initial wave amplitudes $a$.}
	\label{tab:BEs}
	\end{center}
\end{table}
We find that the mass scales roughly like $a^2$, i.e., the energy of the space-time due to the ingoing gravitational wave is approximately quadratic in its amplitude, as one would expect from the linearized theory of gravitational waves.}

\subsection{Quasi-normal ringing}
\label{sec:quasinormal-ringing}

We were also interested in the onset of quasi-normal ringing. In particular, we wanted to see where the ringing is located, and how quickly the decay to the linearised regime occurs. To this end, we directed several modes onto the black hole and watched its reaction~\cite{Frauendiener:2023}.

\begin{figure}[htb]
  \centering
  \includegraphics[width=0.8\textwidth]{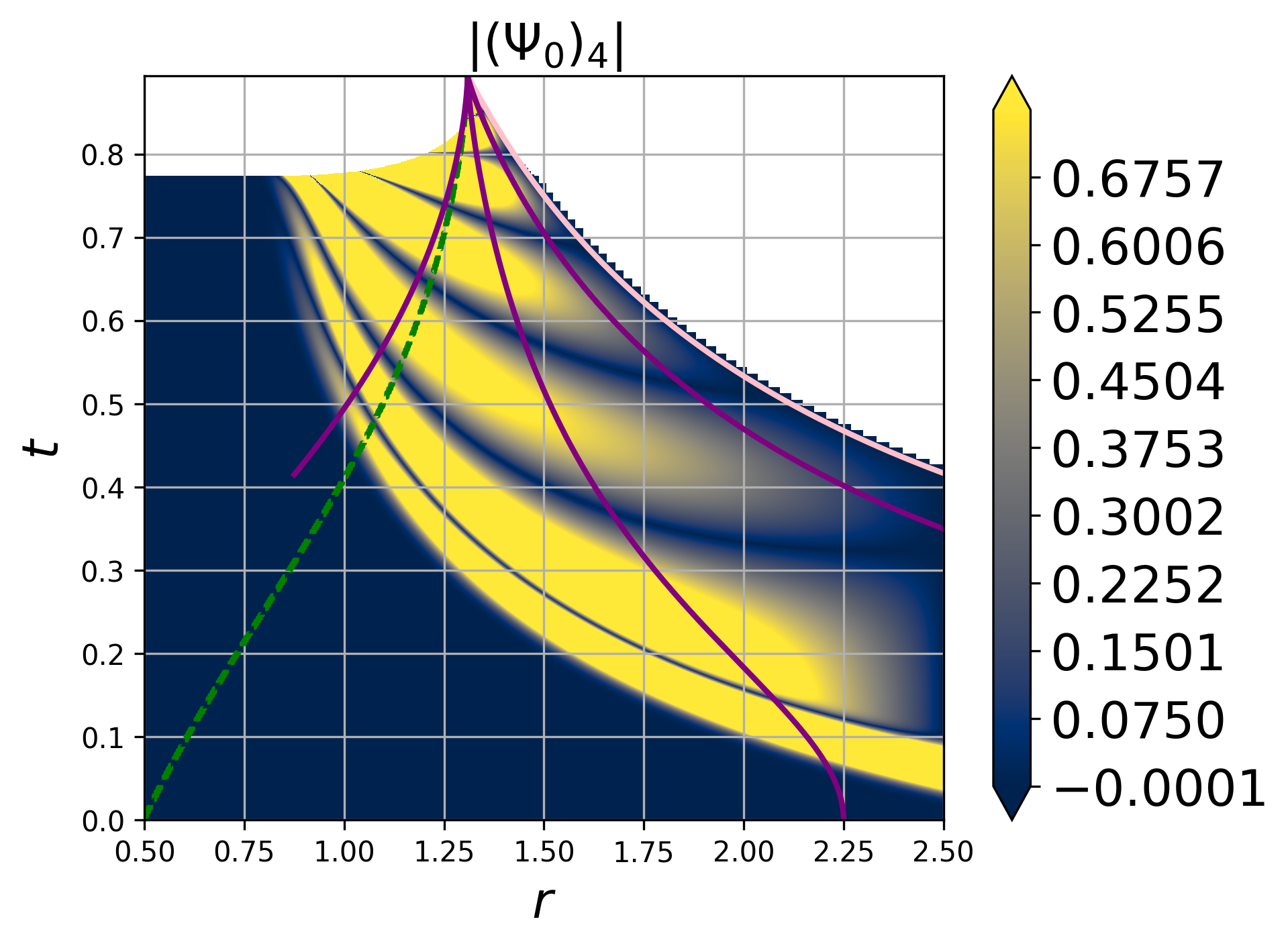}
  \caption{Contour plot of the magnitude of the ingoing $l=4$ mode $\Psi_0$ over the entire computational domain for a fixed angular direction. The green dashed line approximates the horizon. The red lines are quasi-static observers.}
  \label{fig:qnm}
\end{figure}

Fig.~\ref{fig:qnm} shows a contour plot of the magnitude of $\Psi_0$ over the equatorial plane for the $l=4$ mode. The lowest yellow band corresponds to the ingoing pulse. However, it is clearly visible that the ringing starts immediately, creating a back-scattered ingoing pulse, whose intensity increases the further inside one looks. We can see several ``lobes'' corresponding to different ringing phases. The green dashed line in the diagram corresponds to the sequence of largest totally trapped spheres on the hypersurfaces of constant time. In that sense, it approximates the horizon. It is clear that there is nothing special happening at the horizon. 

We also tried to recover the well-known quasi-normal frequencies from this simulation. Again, this turned out to be not entirely straightforward, the reason being that we are trying to compare a linear with a non-linear situation. In the linear case, the frequencies are measured with respect to the preferred time in the Schwarzschild solution with respect to which the black hole is static, i.e., the Killing time. This is also the proper time of a static observer at infinity. Static observers at finite locations suffer time dilation with respect to the asymptotic observer. Instead, in our code we have a global time coordinate determined by the conformal Gauß gauge, which has no obvious relationship with any of the time notions in the linear case.

In order to make a sensible comparison, we introduced quasi-static observers, determined their physical proper time and accounted for the time dilation. Such a quasi-static observer is a world-line which is determined by keeping the angular coordinates and the physical radius fixed. In the absence of radiation these observers become the static observers in the Schwarzschild space-time.

Given the correct time parameter, we trace the field components along the observer's world-line and determine the periods of various $l$-modes by measuring the half-periods, i.e., the times from minimum to minimum. We show some of the results in Table~\ref{tab:frequencies},
\begin{table}[htb]
  \centering
  \small
  \begin{tabular}{ |c|c|c|c|c|c|c| } 
    \hline
    {} & $f_{qnm}$ & $E_1$ & $E_2$ & $E_3$ & $E_4$ & $E_5$ \\ 
    \hline
    $(\Psi_0)_{l=2}$ & 0.58720 & 184.5\% & 53.2\% & 76.4\% & 117.5\% & 73.9\%\\ 
    \hline
    $(\Psi_0)_{l=4}$ & 1.27157 & 2.20\% & 1.10\% & 0.344\% & 0.616\% & 0.156\% \\ 
    \hline
    $(\Psi_4)_{l=2}$ & 0.58720 & 197.4\% & 50.8\% & 90.7\% & 85.7\% & 88.5\% \\ 
    \hline
    $(\Psi_4)_{l=4}$ & 1.27157 & 1.22\% & 0.306\% & 0.412\% & 0.808\% & 0.052\%  \\ 
    \hline
  \end{tabular}
  
  \caption{\label{tab:frequencies}A comparison along the quasi-static world-line at $r\approx3.31$ of the non-linear curvature oscillations to linear theory with an ingoing gravitational pulse proportional to ${}_2Y_{40}$.}
\end{table}
where we list the frequencies $f_{qnm}$ of the quasi-normal modes with $l=2$ and $l=4$ as listed in the literature, see~\cite{Kokkotas:1999} and references therein. The other columns labeled as $E_1$ to $E_5$ list the percentage error between the linear frequency and the ``frequency'' measured along a quasi-static observer (here, located at $r\approx3.31$) between successive minima. We compare both the $l=2$ and $l=4$ modes in the non-linear response in $\Psi_0$ and $\Psi_4$ to a pure $l=4$ ingoing pulse.

Two aspects are clearly visible: first, the excited mode, here $l=4$, behaves very similar to the linear case, and second, the agreement gets better the later the comparison is made. This means that there is a brief transient behaviour in the response which depends on the specifics of the ingoing pulse and dies away after some time. It also shows that the induced ``foreign'' mode (here, the $l=2$ mode) is not the same as the linear response to an $l=2$ excitation, or that it takes much longer to decay to the linear regime. This is due to the fact, that this mode is a secondary phenomenon, created by the non-linear self-interaction.

\subsection{Newman-Penrose constants}
\label{sec:newm-penr-const}

An interesting feature of relativistic ``asymptopia'' are the Newman-Penrose (NP) constants~\cite{Newman:1968}. In our context, these are five complex quantities related to the rescaled Weyl curvature which are defined in terms of an integral over a cut of $\scri^+$. They are constant in the sense that their defining integral evaluates to the same numbers irrespective of which cut is used. This constancy is in contrast to a balance law, as obeyed by the energy-momentum at infinity, which gives the difference between the values of a quantity on two different cuts in terms of a flux integral between them. 

The common interpretation of these quantities is that they are the ``ideal'' value of the rescaled Weyl tensor at time-like infinity $i^+$ even when this point is singular and the space-time is not well-defined there. This view stems from the close relationship between the defining integral and the d'Adh\'emar integral~\cite{Penrose:1960} (similar to the Kirchhoff integral in Maxwell theory) which expresses the value of a field at a point in terms of an integral over a spherical cross-section of an initial or characteristic data surface

From a mathematical point of view the NP constants are interesting because their property of being constant is tied to the smoothness of $\scri$ and, therefore, relates back to the discussion in the introduction. If null-infinity is not sufficiently smooth, then there can be terms which generate a time-dependence of the NP constants. The explicit conditions that have to be satisfied on a cut are spelled out in~\cite{Andersson:1992,Andersson:1994,Chrusciel:1997a}.

\begin{figure}[htb]
\centering
\includegraphics[width=0.8\textwidth]{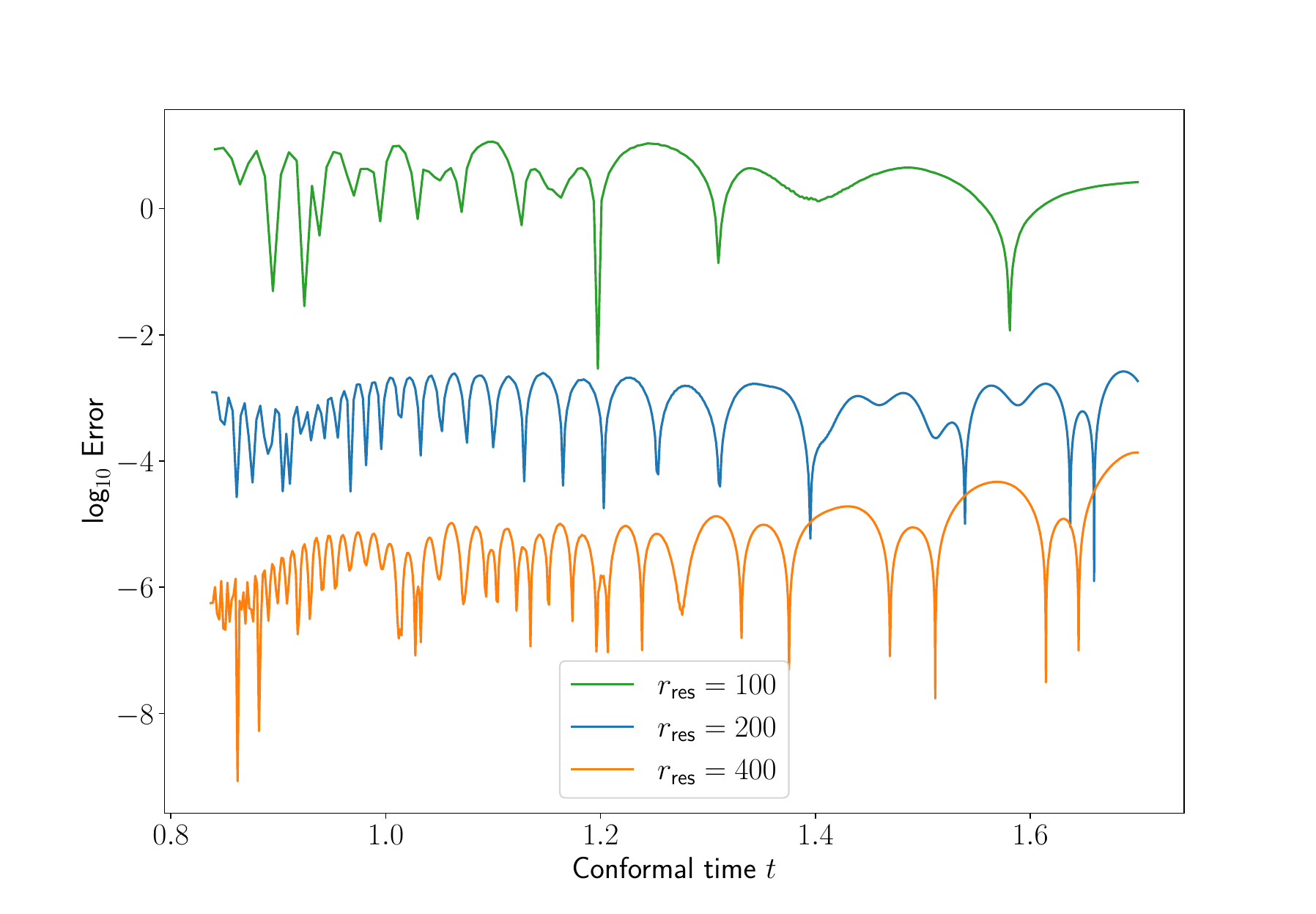}
\caption{Convergence of the log10 difference between the magnitude of the NP constant at time t and at the initial cut with increasing spatial resolution. \new{This figure was taken from~\cite{Frauendiener:2024b}.}}
\label{fig:NPconstants}
\end{figure}

This raises an interesting question. Given that numerical solutions are rarely exact, it is conceivable that numerical errors trigger "non-smooth" modes with the consequence that the NP quantities can not remain constant. We have checked the behaviour of the NP quantities with our numerical simulations~\cite{Frauendiener:2024b} and found that, within numerical accuracy, they are constant, see Fig.~\ref{fig:NPconstants}. In this simulation of an axisymmetric situation, we looked at the value of the only non-vanishing NP constants for successive cuts along $\scri^+$.

How is this finding to be interpreted? There are two possibilities. The obvious and most probable answer is that the violation of smoothness is still beyond the level of accuracy. However, it might also be that the smooth structure on $\scri$ is stable under perturbations due to non-smooth terms. This would be an interesting result which deserves some independent attention. 

The NP constants, which exist on both future and past null-infinity, have gained some interest recently also from another point of view. Penrose~\cite{Penrose:2024} suggested that they could be interpreted as the advanced and retarded fields of a particle whose gravitational field is idealised as an isolated system. This is a bold conjecture, which is not easy to verify and it is not clear what the consequences of this property could be if it would turn out to be true.

\subsection{Further questions}
\label{sec:further-questions}

The framework that we presented here can be used to study several other questions. One could, for instance, ask whether one can ``kick'' the black hole, i.e., what would happen if the gravitational radiation comes in predominantly from one direction? This is different to the situations which we have studied so far in the sense that the ingoing radiation did not carry any linear momentum. In the suggested case, we expect to see a response which carries away linear momentum from the system and we should be able to compute the linear Bondi-Sachs momentum in the emitted radiation. While the linear momentum on a cut of $\scri^+$ is a well-defined quantity, this is not the case when one considers any other finite spherical 2-surface. The quasi-local notion of energy-momentum is not a universally agreed upon unambiguously defined quantity, and there are several different definitions for it, see~\cite{Szabados:2009}. Therefore, it will be difficult to obtain a quantitative relationship between the ingoing and outgoing momenta. However, it might be interesting to explore the cross-section of the scattering process between gravitational waves and the black hole.

A similar question is as to whether we can spin up the black hole? Is it possible to inject gravitational radiation which carries angular momentum into the black hole, thereby spinning it up? This would create a rotating excited black hole which might then settle down to a Kerr black hole. This situation suffers from the same deficiency as the previous one, namely that the notion of angular momentum on the quasi-local level is not unambiguously defined. In this situation, the problem at infinity is also not entirely resolved even though there seems to be a reasonably well-defined notion of angular momentum given by Dray and Streubel~\cite{Dray:1984}.

Of course, this numerical setup can be applied to other black holes also. In principle, switching to a Reissner-Nordstr\"om or, even more general, a Kerr-Newman black hole should consist only in setting up the appropriate initial data obtained from the exact solution. A complication arises in determining the corresponding initial data for the conformal Gauß gauge but this should be straightforward to resolve.

\section{Conclusion}
\label{sec:conclusions}

In this paper, we have presented the current state of the art in our efforts to use the GCFE for numerical evolutions in GR. We have discussed the origin of the equations, the conformal Gauß gauge as the main ingredient, our numerical implementations, and finally, the applications that we have developed so far.

Our discussions show that with our code we can, for the first time, give a clear relationship between initial data, boundary data, and the asymptotic quantities generated on $\scri^+$. We have demonstrated that this approach can be used to analyse the physical quantities at null-infinity. We have verified that the Bondi-Sachs mass
loss formula is valid to a remarkably accurate degree. We have shown that the quasi-normal ringing does not originate at the horizon of a black hole but that it is a reaction of the global fabric of the space-time to the deformation by the incoming gravitational radiation. Finally, we have shown that the NP quantities are constant within the numerical accuracy of our computation.

All of these achievements are possible due to the single most important ingredient, the conformal Gauß gauge. It is very fortunate that the congruence of time-like conformal geodesics which is the basis of this gauge covers the entire Kruskal extension of the Schwarzschild solution. This allows us to carry the simulation up to and beyond $\scri$ and gives us full access to the asymptotic quantities. The CGG has the additional effect of simplifying the equations drastically. Other formulations of the evolution equations~\cite{Frauendiener:1998,Hubner:1996a} contain many more wave type subsystems, which translates into much more computational effort being spent on the boundaries.

But all this comes at a price. There is almost no freedom in the choice of the conformal Gauß gauge, which implies that there is no possibility to adapt the gauge to local variations in the fields. The rigidity of the CGG has the drawback that there is almost nothing that could be done to avoid the rapid ``compactification'' of the computational domain. The fact that we can reach null-infinity in a finite time means that small time-steps in conformal (code) time translate into increasingly large physical time intervals. In a discretised setup, like any computation with a machine epsilon (such as IEEE arithmetic, every conformal time-step will ultimately correspond to an infinite physical time interval. Or, what is the same thing, any fixed physical time-step will ultimately lead to underflow because the corresponding conformal time-step can no longer be represented as a floating point number anymore.

This is a problem of a principal nature. We have tried to mitigate this issue (see our discussion in~\cite{Frauendiener:2023}) but only with minor success. We tried to use the rescaling freedom in the parametrisation of the conformal geodesics but this only delays the problem for a bit. Referring back to Fig.~\ref{fig:qnm}, we see the consequence of this property. The ringing modes are all squeezed into the small region below the singular point~$i^+$. This is to some extent an artefact of how we produced the image because it is drawn with respect to the numerical coordinates $t$ and $r$. In these coordinates the computational domain collapses down to, ultimately, a single radius. Due to dynamical rescaling of the coordinate range we maintain roughly the same number of discrete $r$-values throughout the simulation\footnote{This accounts for the steps outside of $\scri^+$ visible in the figure.}.

However, the fundamental problem concerning the time-steps is not affected by this. Even if we would draw the diagram so that the domain does not peak so rapidly towards $i^+$ we would not see more ``rings'' because the stretching of the time coordinate wipes out the resolution that is necessary to resolve the individual ``rings''. In addition to this there is the obvious fact that the simulation approaches a singularity so that all the relevant quantities blow up.

It would be very useful if one could find a way to slow down the rapid approach to $i^+$. This would certainly require the CGG to be altered but maybe it is possible to find some trade-off which would allow us to maintain most of the advantages of this gauge. Work on this problem is ongoing.

\sloppy
\printbibliography

\end{document}